\documentclass[onecolumn,prd,amsmath,amssymb,floatfix,10pt,superscriptaddress,nofootinbib,eqsecnum,longbibliography]{revtex4-2}

\usepackage{graphicx}
\usepackage{dcolumn}
\usepackage{bm}
\usepackage{hyperref}
\usepackage{xcolor}
\usepackage{natbib}

\usepackage[normalem]{ulem}

\hypersetup{
    colorlinks=true,
    linkcolor=blue,
    filecolor=magenta,      
    urlcolor=blue,
    citecolor=blue
}

\makeatletter
\newcommand\footnoteref[1]{\protected@xdef\@thefnmark{\ref{#1}}\@footnotemark}
\makeatother

\begin{document}
    
    %\title{Constraining the neutron-star pressure anisotropy with gravitational-wave data}
\title{Gravitational-Wave Constraints on Neutron-Star Pressure Anisotropy \\ via Universal Relations}
    
    \author{Victor Guedes}
    \email{tpx5df@virginia.edu}
    \affiliation{Department of Physics, University of Virginia, Charlottesville, VA 22904, USA}

    \author{Siddarth Ajith}
    \affiliation{Department of Physics, University of Virginia, Charlottesville, VA 22904, USA}

    \author{Shu Yan Lau}
    \affiliation{Department of Physics, Montana State University, Bozeman, MT 59717, USA}

    \author{Kent Yagi}
    \affiliation{Department of Physics, University of Virginia, Charlottesville, VA 22904, USA}
    
    \begin{abstract}

        Neutron stars may exhibit pressure anisotropy arising from various physical mechanisms, such as elasticity, magnetic fields, viscosity, and superfluidity. We compute the tidal deformability and the $f$-mode oscillation frequency of anisotropic neutron stars using a phenomenological quasi-local model characterized by a single dimensionless anisotropy parameter. We find that while the relation between the tidal deformability and the $f$-mode frequency depends on the degree of anisotropy, it remains largely insensitive to variations in the equation of state (the relation between radial pressure and energy density) for a fixed anisotropy parameter, similar to the isotropic case. Leveraging this anisotropy-dependent universal relation within a statistical framework, we place constraints on the anisotropy parameter using both the gravitational-wave observation of GW170817 and simulated data for a GW170817-like event observed by a future network of detectors. We find that the anisotropy parameter can be constrained to order unity with current data, and the bounds remain comparable with future detector sensitivities. Importantly, these constraints are only weakly affected by uncertainties in the neutron-star equation of state.

    \end{abstract}
    
    \maketitle
    
    \section{Introduction}

        The theoretical modeling of neutron stars (NSs) relies on the description of matter at extreme densities, in regimes that are difficult to access by terrestrial experiments. In particular, the relation between the pressure and the energy density of cold, catalyzed matter in the interior of NSs, commonly referred to as the equation of state (EOS), is necessary to compute macroscopic NS observables. Recent measurements of masses of high-mass pulsars from radio observations~\cite{Antoniadis:2013pzd,Fonseca:2016tux,NANOGrav:2019jur,Romani:2022jhd,Saffer:2024tlb}, joint mass-radius measurements of pulsars from X-ray observations~\cite{Miller:2019cac,Riley:2019yda,Miller:2021qha,Riley:2021pdl}, and tidal deformabilities of NSs in binary systems from gravitational-wave (GW) observations~\cite{LIGOScientific:2017vwq,LIGOScientific:2020aai} have put significant bounds on the pressure-density relation. Nevertheless, despite the variety of available information, from astrophysical observations to nuclear theory and experiments~\cite{Koehn:2024set}, uncertainties in the high-density regime ($\sim10^{14}-10^{15}$ g/cm$^{3}$) still persist.

        GW observations of inspiralling NSs can constrain the EOS through the measurement of the NS tidal deformability. This observable quantifies how strongly one star responds to the tidal field of its companion during the early stages of the inspiral, when variations in the tidal field can be treated adiabatically. While the individual tidal deformabilities of the stars in GW170817 could not be inferred as precisely as the effective tidal deformability, which is a mass-weighted combination of the two, the existing bounds could still be used to rule out overly stiff EOSs~\cite{LIGOScientific:2018cki}.
        
        Another important EOS-sensitive quantity that enters in the gravitational waveform is the frequency of the fundamental oscillation mode (or the $f$-mode) of the stars. The $f$-mode of inspiralling NSs can leave signatures in the late-inspiral GW signal, when variations in the tidal fields resonate with the frequency of the $f$-mode, i.e. when tidal fields become dynamical. Current observational bounds on the $f$-mode due to dynamical tides are weak because of its high frequency ($\sim 2-4$ kHz), where current ground-based detectors are not so sensitive. The analysis in Pratten et al.~\cite{Pratten:2019sed}, for example, rules out anomalously low values for the frequency but does not provide an upper limit without further crucial assumptions. Next-generation detectors, such as Cosmic Explorer (CE) and Einstein Telescope (ET), are expected to provide significantly tighter constraints on the $f$-mode frequency.

        NS observables cannot be precisely predicted given the current uncertainties in the EOS. Nevertheless, certain relations between dimensionless NS quantities have a weak EOS dependence, as shown by e.g. Yagi \& Yunes~\cite{Yagi:2013bca,Yagi:2013awa}, for the relation between the moment of inertia, spin-induced quadrupole moment, and tidal deformability (the ``I-Love-Q'' relations) and by Chan et al.~\cite{Chan:2014kua} for the relation between the $f$-mode and the tidal deformability (the ``$f$-Love'' relation); see~\cite{Yagi:2016bkt} for a comprehensive review on this topic. These are usually referred to as (quasi-) universal relations (URs), and have been shown to hold for several EOS models. The $f$-Love relation is of particular interest, because both the tidal deformability and the $f$-mode frequency affect GWs from binary NS mergers. Thus, given a joint inference of the parameters in this relation, we can probe alternative theories of gravity~\cite{Yagi:2016bkt,Doneva:2017jop,Gupta:2017vsl,Silva:2020acr,Saffer:2021gak}, exotic compact objects~\cite{Pani:2015tga,Uchikata:2016qku,Adam:2022nlq,Adam:2024zqr}, or, in our case, the amount of pressure anisotropy in NSs~\cite{Yagi:2015hda}. 
        %In particular, we highlight the current bounds reported in Pratten et al.~\cite{Pratten:2019sed} and prospective bounds for third generation GW detectors reported in Williams et al.~\cite{Williams:2022vct}.
        
        Typically, EOS models for nuclear matter are used to describe a spherically symmetric, isotropic fluid distribution, in which the pressure is the same in the radial and tangential directions. This stellar model is then perturbed accordingly to compute observable quantities such as the tidal deformability or the $f$-mode frequency of NSs. However, local anisotropy can arise in NS interiors due to e.g. strong magnetic fields~\cite{Yazadjiev:2011ks}, superfluidity~\cite{Herrera:1997plx}, viscosity~\cite{1992Ap&SS.193..201B}, or elasticity~\cite{Karlovini:2002fc}. The pressure anisotropy can affect NS observables in a significant way, and can pose alternative interpretations for current and future observations.

        Attempts to model the NS pressure anisotropy have been mostly phenomenological, e.g. Bowers \& Liang~\cite{Bowers:1974tgi} and Horvat et al.~\cite{Horvat:2010xf}. The use of such models is justified by the relatively unconstrained nature of anisotropy in NSs. Several studies have reported calculations of NS observables considering such phenomenological models: Silva et al.~\cite{Silva:2014fca} computed the moment of inertia while Yagi \& Yunes~\cite{Yagi:2015hda} studied the spin-induced quadrupole moment and tidal deformability; Doneva et al.~\cite{Doneva:2012rd} computed the $f$-mode within the Cowling approximation by ignoring spacetime perturbations, which has recently been extended to full general relativity by Mondal and Bagchi~\cite{Mondal:2023wwo,Mondal:2025ixk}, Lau et al.~\cite{Lau:2024oik}, and Arbanil et al.~\cite{Arbanil:2025jep} (we point out some potential issues in~\cite{Mondal:2023wwo,Mondal:2025ixk} in Appendix~\ref{ap_f}). The measurement of such quantities can place bounds on the anisotropy of NSs and such bounds can be used to guide the theoretical development of physically motivated models. 
        %For example, tidal deformability measurements with GW observations have been used to place bounds on anisotropy~\cite{} \ky{Please check Refs.~[5-21] in my note and cite relevant papers here.}. 

    %    One of the immediate and most advocated applications of URs is the inference of NS quantities that cannot be directly measured. 
        Traditionally, astrophysical observations and experimental nuclear physics measurements can be used to constrain the EOS in a Bayesian fashion, as in, e.g.~\cite{2020ApJ...888...12M,Koehn:2024set}. Usually, the posterior on EOS parameters are obtained from a large set of phenomenological EOS models used to compute the prior and the likelihood of the data, given the parameters of these models. In the case of phenomenological anisotropic NSs, there is an extra parameter that determines the degree of anisotropy, which can also be inferred, as recently done by Pang et al.~\cite{Pang:2025fes} (see also~\cite{Biswas:2019gkw,Rahmansyah:2020gar,Das:2020zbf,2022EPJC...82..136D,Arbanil:2023yil,Arbanil:2021ahh,Das:2021qaq,Das:2022ell,Pretel:2023nlr,Pretel:2023nhf,Mohanty:2023hha,Das:2023qej,Lopes:2024wfk,Mahapatra:2024ywx,Pretel:2024pem,Jyothilakshmi:2024zqn,Liu:2025cwy} for related works on constraining pressure anisotropy through NS observations, in particular, tidal deformability measurement with GWs). However, as we show in this work, we can approach this inference problem using URs.
        
        In this work, we first study how the $f$-Love relation changes when we consider pressure anisotropy. We use a phenomenological model inspired by Horvat et al.~\cite{Horvat:2010xf}, which relies on a single dimensionless parameter, to compute the tidal deformability and $f$-mode frequency for NSs. We find that the $f$-Love relation depends on such anisotropy parameter, though it remains EOS-independent (i.e., for different choices of the radial pressure vs. energy density relation) for a fixed anisotropy parameter. 
        
        We next demonstrate how one can use the anisotropy-dependent but EOS-independent $f$-Love relation to infer the pressure anisotropy in NSs through GW observations of binary NS mergers.
        %We also provide a constraint on the dimensionless anisotropy parameter considering the first GW detection from a binary NS. 
        More specifically, we use the inference of the tidal deformability and $f$-mode frequency for GW170817 in Pratten et al.~\cite{Pratten:2019sed}. We also study future prospects, using the results for a simulated GW170817-like event for a network of third-generation GW detectors in Williams et al.~\cite{Williams:2022vct}. 
        %We use these anisotropy-dependent, EOS-insensitive $f$-Love relations in our statistical approach.  
        We find that, with both current and simulated data for future detectors, the anisotropy parameter can be constrained to order unity.
        
        The rest of the paper is organized as follows. In Sec.~\ref{sec2}, we describe the phenomenological anisotropy model for NSs studied in this paper and study the $f$-Love relation for such stars. In Sec.~\ref{sec3}, we describe our statistical approach; show constraints on the parameter that controls the NS pressure anisotropy, using the joint inference for GW170817 in Pratten et al.~\cite{Pratten:2019sed}; and explore future prospects using the simulation results in Williams et al.~\cite{Williams:2022vct}. In Sec.~\ref{sec4}, we summarize our findings and provide possible avenues for future work. We use the geometric units of $c=G=1$ throughout the paper.

    \section{The \textbf{\textit{f}}-Love relation for anisotropic neutron stars}
    \label{sec2}

        In this Section, we report how the $f$-Love relation between the tidal deformability and the $f$-mode frequency changes for anisotropic NSs from the isotropic case. We first describe the phenomenological anisotropy model adopted in this paper, that was used to study radial stability of anisotropic NSs in Appendix~\ref{apA}. We then present key definitions for the tidal deformability and the fundamental mode of NSs. The details on how to compute these observables for anisotropic NSs have already been published in the previous literature~\cite{Yagi:2015hda,Lau:2024oik}. 
        %Yagi \& Yunes~\cite{Yagi:2015hda} derived the perturbation equations for tidal deformations and Lau et al.~\cite{Lau:2024oik} reported the equations for non-radial oscillations. 
        We review the formalism for both cases in Appendix~\ref{ap_Lf}.

        \subsection{Pressure Anisotropy}
        \label{subsec_aniso}
        
            The NS pressure anisotropy is defined as
            \begin{align}
                \sigma=p_{r}-p_{t},
            \end{align}
            where $p_{r}$ is the radial pressure and $p_{t}$ is the tangential pressure. In this paper, we adopt the model inspired by the one in Horvat et al.~\cite{Horvat:2010xf} and is given in terms of $p_{r}$ and the quasi-local variable $\mu=2m/r$, where $m=m(r)$ is the mass in a sphere of radius $r$:
            \begin{align}
                \sigma=\sigma(p_{r},\mu). \label{ansatz0}
            \end{align}
            For example, from a dimensional analysis, one can consider the following form with dimensionless constants $c_k$:
            \begin{equation}
                \sigma = p_r \sum_{k=0}^{\infty} c_k \, \mu^k\,.
                \label{eq:sigma_series}
            \end{equation}
            This ansatz corresponds to expanding $\sigma$ in terms of a compactness parameter $\mu$~\footnote{The compactness is usually defined as $C=M/R$, where $M$ is the total mass of the star and $R$ is the radius, so $C=\mu(r=R)/2$.} and higher order terms encode relativistic effects. The original model considered in~\cite{Horvat:2010xf} corresponds to setting all the $c_k$ constants to 0 except for $c_1$.
     
            Let us impose two regularity conditions at the stellar center $r=0$:
            \begin{enumerate}
                \item Regularity of the Tolman-Oppenheimer-Volkoff (TOV) equation (see Eq.~\eqref{eqpr} in Appendix~\ref{ap_Lf}) requires 
                \begin{align}
                    \sigma\sim\mathcal{O}(r^{2}) \; \; \; {\rm for} \; \; \; r\rightarrow 0.
                    \label{reg_con_1}
                \end{align}
                Given that $p_{r} \sim \mathcal{O}(\mu^0)$ while $\mu \sim \mathcal{O}(r^2)$, the above condition forces $c_0 = 0$.
                \item Regularity of the equation for the Lagrangian perturbation of $p_{r}$ (see Eq.~\eqref{eqX} in Appendix~\ref{ap_f}) requires
                \begin{align}
                    \left(\frac{\partial{\sigma}}{\partial{p_{r}}}\right)_{\mu},\left(\frac{\partial{\sigma}}{\partial{\mu}}\right)_{p_{r}}\sim\mathcal{O}(r^{2})\; \; \; {\rm for} \; \; \; r\rightarrow 0. \label{reg_con_2}
                \end{align}
                This further forces $c_1 = 0$.
            \end{enumerate}
            In this paper, we adopt the simplest form of Eq.~\eqref{eq:sigma_series} satisfying the above two conditions:
            \begin{align}
                \sigma=\beta p_{r}\mu^{2},
                \label{ansatz}
            \end{align}
            where $\beta$ is the free parameter that controls the anisotropy. Note that there is an extra factor of $\mu$ compared to the one in~\cite{Horvat:2010xf}. Equation~\eqref{ansatz} is also adopted in~\cite{Lau:2024oik,Becerra:2025wno}. Interestingly, the model in Eq.~\eqref{ansatz} can be mapped to the post-TOV framework in~\cite{Glampedakis:2015sua} up to the second post-Newtonian order that was developed to capture beyond-Einsteinian effects in the TOV equations\footnote{More specifically, the model in Eq.~\eqref{ansatz} can be mapped to the F3 family term $mp_r/r\rho_{b}$, where $\rho_{b}$ is the baryonic density, with constant $\pi_{3}=8\beta$.}. 

%%%%%%%%
            
            We only consider physically-viable stars by imposing the following conditions to the background, spherically-symmetric solutions {\color{black}(see also~\cite{Suarez-Urango:2023ikq, Becerra:2024wku})}:
            % \begin{enumerate}
            %     \item weak energy condition:
            %     \item null energy condition:
            %     \item strong energy condition:
            %     \item dominant energy condition:
            %     \item positivity of pressure:
            %     \item causality: $$
            % \end{enumerate}
            %
            % When solving the background equations (see Eqs.~\eqref{eqm}$-$\eqref{eqpr}) and obtaining the equilibrium solutions, we impose the validity of the energy conditions~\cite{Poisson:2009pwt}:
            \begin{align}
                &\textrm{1. weak energy condition~\cite{Poisson:2009pwt}: } \quad \rho\geq0,\quad \rho+p_{r}>0,\quad \rho+p_{t}>0, \label{wec}\\
                &\textrm{2. null energy condition~\cite{Poisson:2009pwt}: }\quad \rho+p_{r}\geq0 \quad \rho+p_{t}\geq0, \label{nec}\\
                &\textrm{3. strong energy condition~\cite{Poisson:2009pwt}: }\quad \rho+p_{r}+2p_{t}\geq0, \quad \rho+p_{r}\geq0, \quad\rho+p_{t}\geq0, \label{sec}\\
                &\textrm{4. dominant energy condition~\cite{Poisson:2009pwt}: }\quad \rho\geq0, \quad \rho\geq|p_{r}|, \quad \rho\geq|p_{t}|, \label{dec} \\
                &\textrm{5. positivity of pressure: }\quad  p_{r}\geq0,  \quad p_{t}\geq0, \\
                &\textrm{6. causality: }\quad  0 \leq c^{2}_{s,r}, c^{2}_{s,t} \leq 1, \label{causality}
            \end{align}
            % besides imposing positivity of the radial and tangential pressure:
            % \begin{align}
            %     p_{r}\geq0\textrm{ and }p_{t}\geq0.
            % \end{align}
            % We also impose the validity of the causality conditions:
            % \begin{align}
            %     0 \leq c^{2}_{s,r},c^{2}_{s,t} \leq 1, \label{causality}
            % \end{align}
            where $\rho$ is the energy density {\color{black} while $c^{2}_{s,r}=(\partial{p_{r}}/\partial{\rho})_{\rm C}$ and $c^{2}_{s,t}=(\partial{p_{t}}/\partial{\rho})_{\rm C}$, where ``C'' is a condition that is applied on the derivative. In equilibrium, these partial derivatives reduce to total derivatives, i.e. 
            \begin{align}
                c^{2}_{s,r}=\left(\frac{{\rm d}p_{r}}{{\rm d}\rho}\right)_{\rm EOS} \quad {\rm and} \quad c^{2}_{s,t}=\left(\frac{{\rm d}p_{t}}{{\rm d}\rho}\right)_{\rm EOS},\label{sound_speed}
            \end{align}
            where the derivative is taken from the $p_{r}$~vs.~$\rho$ and $p_{t}$~vs.~$\rho$ relations.} We further require the radial stability of anisotropic NSs. In Appendix~\ref{apA}, we confirm that NSs with anisotropic pressure described by Eq.~\eqref{ansatz} become radially unstable when $\partial M/\partial \rho_0 < 0$ for the stellar mass $M$ and the central energy density $\rho_0$, similar to the case with isotropic pressure.
            %We categorize any of the models does not satisfy the conditions above in eqs.~\eqref{wec}$-$\eqref{causality}, as unphysical, and thus are not considered in the construction of URs. On top of that, in appendix~\ref{apA}, we also study the radial stability of the models by computing their fundamental radial oscillation mode and confirming that the maximum-mass criterion is valid for anisotropic NSs described by the ansatz in Eq.~\eqref{ansatz}.

        \subsection{Tidal Deformability and \textbf{\textit{f}}-mode Frequency}

            \begin{figure*}
                \centering
                \includegraphics[width=\linewidth]{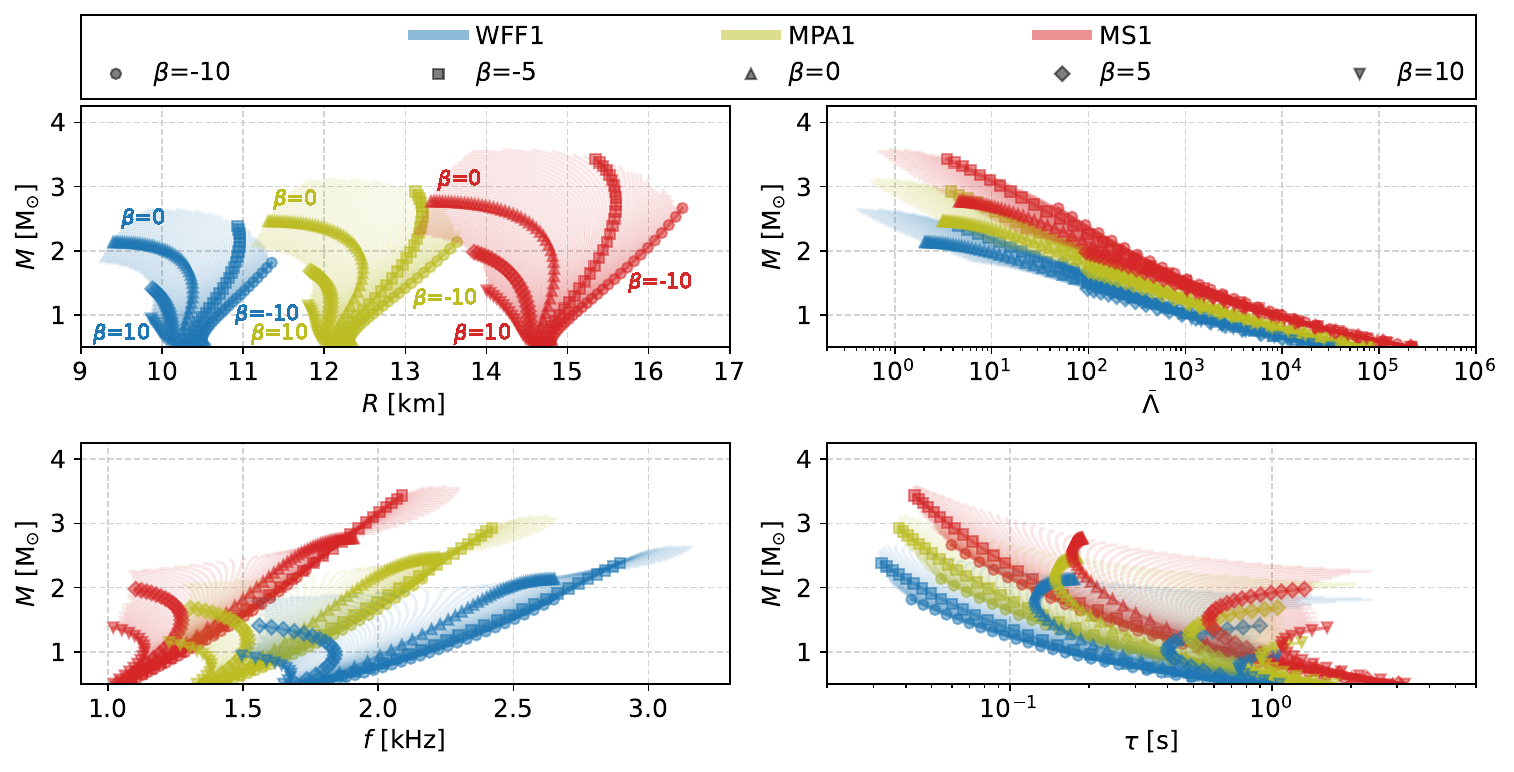}
                \caption{Relation between the mass $M$ and the following quantities: radius $R$ (top left), dimensionless tidal deformability $\bar{\Lambda}$ (top right), $f$-mode frequency $f=\textrm{Re}(\omega)/2\pi$ (bottom left), and $f$-mode damping time $\tau=1/\textrm{Im}(\omega)$ (bottom right). We show results for only three EOSs (WFF1, MPA1, and MS1) and for $\beta\in\left[-10,10\right]$. We highlight results for $\beta\in\{-10,-5,0,5,10\}$ and label curves for $\beta\in\{-10,0,10\}$ in the mass-radius plot, for better readability. We show results for $M>0.5$ M$_{\odot}$ and for physical models, i.e. those that do not violate the conditions in Eqs.~\eqref{wec}$-$\eqref{causality}.}
                \label{fig1}
            \end{figure*}

            The tidal deformability characterizes how easily an object is deformed due to an external tidal field. We focus on the dominant deformation at the quadrupolar order ($\ell =2$) in multipole decomposition. The tidal deformability is then defined by the ratio between the tidally-induced quadrupole moment and the strength of the (electric-type) external tidal field. It is related to the electric-type quadrupolar Love number $k_2$ (see Eq.~\eqref{eqk2}) as
            \begin{align}
                \Lambda=\frac{2}{3}k_{2}R^{5},
            \end{align} 
            %where $k_{2}$ is the $\ell=2$ relativistic electric tidal Love number (see Eq.~\eqref{eqk2}) and 
             where $R$ is the stellar radius. In the pioneering work of Hinderer~\cite{Hinderer:2007mb}, the Love number for NSs with isotropic pressure is obtained by solving the equations for the quadrupolar static perturbations of a spherically symmetric background spacetime and matching the exterior solution (for $r>R$, where $r$ is the coordinate radius) with the asymptotic solution for the gravitational field of a tidally deformed star. We will specifically work with the dimensionless version of the tidal deformability defined as
            \begin{align}
                \bar{\Lambda}=\frac{\Lambda}{M^{5}}. \label{Lambda_bar}
            \end{align}
            For the case with pressure anisotropy, the interior perturbation equations are modified, as originally derived in~\cite{Yagi:2015hda}. In Appendix~\ref{ap_L}, we present the key perturbation equations in a slightly different form. We have checked that they are identical to those in~\cite{Yagi:2015hda}. 
             
            The complex frequency of the fundamental quasi-normal mode, that we refer to as the $f$-mode, is defined as:
            \begin{align}
                \omega=2\pi f+\frac{i}{\tau},
            \end{align}
            where $f$ is the oscillation frequency and $\tau$ is the damping time. The pioneering works of Thorne \& Campolattaro~\cite{1967ApJ...149..591T} and Chandrasekhar \& Ferrari~\cite{1991RSPSA.432..247C} constitute the basis for the study of non-radial dynamical perturbations of relativistic stars with isotropic pressure, especially the $\ell=2$ modes. Lindblom \& Detweiler~\cite{1983ApJS...53...73L, 1985ApJ...292...12D} formulated the problem in the Regge-Wheeler gauge~\cite{Regge:1957td} and obtained simplified perturbation equations. These equations are then solved and matched to an asymptotic outgoing wave solution, which describes an open system that is losing energy through gravitational radiation, and thus determines the quasi-normal mode frequency. Similar to the tidal deformability case, we work with the dimensionless version of the $f$-mode complex frequency defined as
            \begin{align}
                \bar{\omega}=M\omega. \label{omega_bar}
            \end{align}
            For convenience, we also define the real part of Eq.~\eqref{omega_bar} as:
            \begin{align}
                \bar{\Omega}={\rm Re}(\bar{\omega}).
            \end{align}
    
        \subsection{\textbf{\textit{f}}-Love Relation}

            \begin{figure*}
                \centering
                \includegraphics[width=\linewidth]{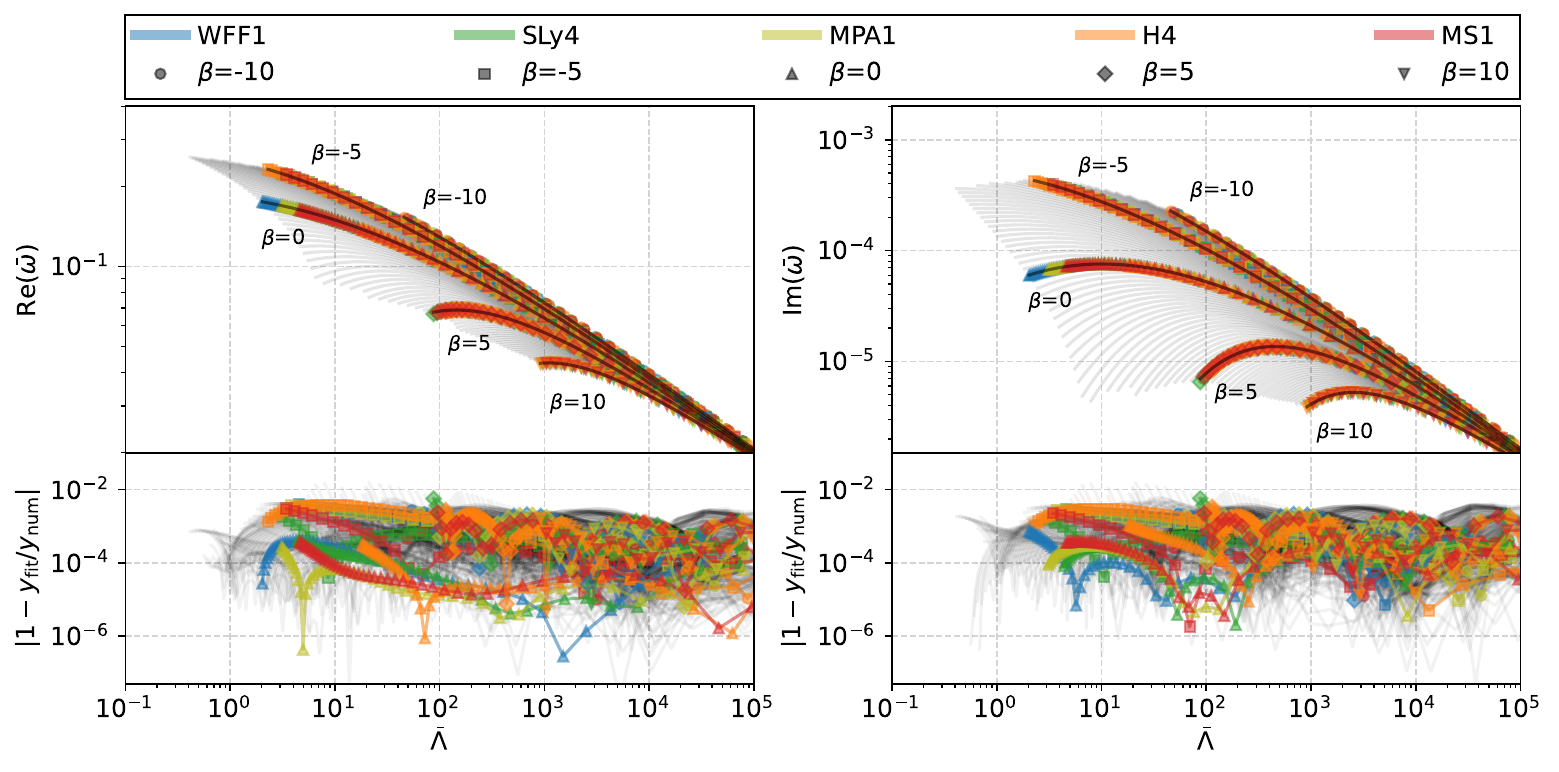}
                \caption{(Top) $f$-Love relation for anisotropic NSs for different choices of the anisotropy parameter $\beta$: the dimensionless real (left panel) and imaginary (right panel) part of the $f$-mode complex frequency (see Eq.~\eqref{omega_bar}) against the dimensionless tidal deformability $\bar{\Lambda}$ (see Eq.~\eqref{Lambda_bar}). (Bottom) The absolute relative error, defined as $|1 - y_{\rm num}/y_{\rm fit}|$, where $y_{\rm num}$ is the numerical result and $y_{\rm fit}$ is the fit result obtained from Eq.~\eqref{fit} for each $\beta$. We show results for five EOSs (WFF1, SLy4, MPA1, H4, and MS1) and for $\beta\in\left[-10,10\right]$. We label curves for $\beta\in\{-10,-5,0,5,10\}$ for better readability. The $f$-Love relation depends on $\beta$ and it remains EOS-independent for a fixed $\beta$.}
                \label{fig2}
            \end{figure*}
        
            Let us now present relations among various NS observables. We consider the following nuclear physics EOSs for the relation between the radial pressure $p_{r}$ and the energy density $\rho$: WFF1~\cite{1988PhRvC..38.1010W}, SLy4~\cite{Douchin:2001sv}, MPA1~\cite{1987PhLB..199..469M}, H4~\cite{Lackey:2005tk}, and MS1~\cite{1996NuPhA.606..508M}. These represent soft, soft-intermediate, intermediate, intermediate-stiff, and stiff EOSs respectively, and cover a wide range of the radius for a NS with 1.4 M$_\odot$, namely $10\textrm{ km} \lesssim R_{1.4} \lesssim 15\textrm{ km}$. 
            
            Figure~\ref{fig1} shows the mass $M$ as a function of the radius $R$ (top left), the dimensionless tidal deformability $\bar{\Lambda}$ (top right),  the $f$-mode frequency $f=\textrm{Re}(\omega)/2\pi$ (bottom left), and the $f$-mode damping time $\tau=1/\textrm{Im}(\omega)$ (bottom right). We show results for only three EOSs (WFF1, MPA1, and MS1) and for $\beta\in\left[-10,10\right]$, although we highlight results for $\beta\in\{-10,-5,0,5,10\}$ and label curves for $\beta\in\{-10,0,10\}$ in the mass-radius plot, for better readability. We only display results for $M \geq 0.5$ M$_{\odot}$ and for physically-viable models, i.e. those that satisfy the conditions in Eqs.~\eqref{wec}$-$\eqref{causality}. Consequently, each curve for non-zero $\beta$ does not extend to the maximum mass. Note that, in general, we have $M(\rho_{0},\beta)<M(\rho_{0},\beta=0)$ for positive $\beta$ and vice-versa for negative $\beta$. We can see similar trends for $\bar{\Lambda}$, $f$, and $\tau$. Besides that, $f$ can take low values ($\sim1-1.5$ kHz) for largely positive $\beta$ for masses in the range $\sim1-1.5$ M$_{\odot}$. Thus, based on the analysis of Pratten et al.~\cite{Pratten:2019sed} for GW170817, we expect that largely positive values for $\beta$ are disfavored, given the low posterior support at low frequencies. This is indeed what we find in Sec.~\ref{sec3}.
            
            We next study the universal $f$-Love relations for anisotropic NSs. Figure~\ref{fig2} presents such relations between the dimensionless real and imaginary parts of the $f$-mode complex frequency (see Eq.~\eqref{omega_bar}) and the dimensionless tidal deformability $\bar{\Lambda}$ (see Eq.~\eqref{Lambda_bar}) for various $\beta$ and EOSs. Remarkably, although the $f$-Love relation depends sensitively on the anisotropy parameter $\beta$, it remains universal for a fixed $\beta$ with respect to the choice of EOS models for the $p_{r}(\rho)$ relation. 
            %We show results for five representative values for $\beta$. We once again note that Re$(\bar{\omega})$ can reach small values ($\sim 0.04$) for high values of $\bar{\Lambda}$ ($\sim10^{3}$), thus reinforcing that large positive $\beta$ is probably not favored by GW170817 data. 
            
            We also show, in the left panel of Fig.~\ref{fig2}, fits for the EOS-insensitive $f$-Love relations (for the real part of the frequency) given by
             \begin{align}
                {y}_{\rm UR}(x,\beta)=\sum^{4}_{i=0}a_{i}(\beta){x}^{i}, \label{fit}
            \end{align} 
            with 
            \begin{equation}
                 x={\rm log}_{10}(\bar{\Lambda}), \quad y={\rm log}_{10}(\bar{\Omega}).
            \end{equation}
            Note that the fitting coefficients depend on $\beta$. We also evaluate the variance as
            \begin{align}
                s_{\rm UR}^{2}(\beta)=\frac{1}{N-1}\sum_{i=1}^{N}[y_{i}-{y}_{{\rm UR}}({x}_{i},\beta)]^{2}, \label{var}
            \end{align}
            where $N$ is the number of data points for $x$. Note that the variance in Eq.~\eqref{var} encodes the EOS variation of the URs. In the top half of Table~\ref{tab1}, we provide the fitting coefficients for the UR fit, their standard deviation, and the average and maximum absolute relative errors for the five values of the anisotropy parameter $\beta$ and five EOSs shown in Fig.~\ref{fig2}. We repeat the same analysis for the $f$-Love relation involving the imaginary part of the frequency (or the damping time) and present the fits and the coefficients in the right panel of Fig.~\ref{fig2} and the bottom half of Table~\ref{tab1}, respectively.

            %\victor{The fact that the $f$-Love relation has a weak EOS dependence but a strong dependence on the anisotropy allows us to establish bounds on the anisotropy parameter in an EOS-independent way, given measurements of the tidal deformability and the $f$-mode frequency. The reason for this anisotropy-induced non-universality is currently unknown. We point out that the pressure anisotropy gives rise to an extra term in the TOV equation (see Eq.~\eqref{eqpr} in Appendix~\ref{ap_Lf}) that acts like an additional force, which may not be captured by simply changing the EOS, although a more in-depth study is needed. Such a study may further provide hints about the origin of the EOS universality, which is also not yet fully understood.} \ky{This paragraph includes some speculation, so I think it'd be better to move it to the discussion section.}
 
    \section{Constraints from GW170817 and future prospects}
    \label{sec3}

        \begin{table*}
            \centering
            \begin{ruledtabular}
                \begin{tabular}{cccccccccc}
                    $\beta$ & $a_{0}$ & $a_{1}$ & $a_{2}$ & $a_{3}$ & $a_{4}$ & $s_{\rm UR}$ & avg($\Delta{y}$) & max($\Delta{y}$)\\
                    \hline
                    $-$10 & $-$0.5656 & $-$0.07258 & $-$0.05770 & 7.286$\times10^{-3}$ & $-$3.594$\times10^{-4}$ & 8.558$\times10^{-4}$ & 6.696$\times10^{-4}$ & 2.752$\times10^{-3}$ \\
                    $-$5  & $-$0.5982 & $-$0.08249 & $-$0.05115 & 6.500$\times10^{-3}$ & $-$3.440$\times10^{-4}$ & 1.264$\times10^{-3}$ & 1.201$\times10^{-3}$ & 3.894$\times10^{-3}$ \\
                    0   & $-$0.7404 & $-$0.03745 & $-$0.05166 & 5.344$\times10^{-3}$ & $-$2.375$\times10^{-4}$ & 1.352$\times10^{-4}$ & 1.165$\times10^{-4}$ & 4.058$\times10^{-4}$ \\
                    5   & $-$2.52   & 1.608     & $-$0.6457  & 0.1022  & $-$6.170$\times10^{-3}$ & 1.606$\times10^{-3}$ & 1.046$\times10^{-3}$ & 5.631$\times10^{-3}$ \\
                    10  & $-$6.479  & 4.816     & $-$1.626   & 0.2351  & $-$0.01289 & 8.103$\times10^{-4}$ & 4.451$\times10^{-4}$ & 1.627$\times10^{-3}$ \\
                    \hline
                    $-$10 & $-$2.975  & $-$0.2542   & $-$0.1062  & 0.01134 & $-$4.865$\times10^{-4}$ & 3.017$\times10^{-3}$ & 5.828$\times10^{-4}$ & 2.200$\times10^{-3}$ \\
                    $-$5  & $-$3.297  & $-$0.1598   & $-$0.1145  & 0.012   & $-$5.705$\times10^{-4}$ & 4.817$\times10^{-3}$ & 9.638$\times10^{-4}$ & 2.957$\times10^{-3}$ \\
                    0   & $-$4.338  & 0.4639    & $-$0.2798  & 0.03489 & $-$1.909$\times10^{-3}$ & 1.395$\times10^{-3}$ & 2.426$\times10^{-4}$ & 1.098$\times10^{-3}$ \\
                    5   & $-$12.54  & 8.057     & $-$3.01    & 0.477   & $-$0.02880 & 7.239$\times10^{-3}$ & 1.176$\times10^{-3}$ & 5.714$\times10^{-3}$ \\
                    10  & $-$29.13  & 21.43     & $-$7.068   & 1.022   & $-$0.05605 & 3.348$\times10^{-3}$ & 4.915$\times10^{-4}$ & 1.797$\times10^{-3}$
                    \end{tabular}
                \caption{Coefficients for the $f$-Love URs for anisotropic NSs, using the fitting function in Eq.~\eqref{fit}. We show results for five values of $\beta$ for log$_{10}$(Re($\bar{\omega}$)) vs. log$_{10}$($\bar{\Lambda}$) relation (top half, corresponding to the left panel of Fig.~\ref{fig2}) and for log$_{10}$(Im($\bar{\omega}$)) vs. log$_{10}$($\bar{\Lambda}$) relation (bottom half, corresponding to the right panel of Fig.~\ref{fig2}). The last three columns display the standard deviation, defined as the square root of the variance in Eq.~\eqref{var}, the average, and the maximum for the absolute relative error, defined as $\Delta{y}=|1-y_{\rm fit}/y_{\rm num}|$, for the URs.}
                \label{tab1}
            \end{ruledtabular}
        \end{table*}

        In this Section, we use the anisotropy-dependent, EOS-independent $f$-Love relations to infer the anisotropy parameter $\beta$, given inferences of $x$ and $y$ from GW data. Pratten et al.~\cite{Pratten:2019sed} obtained the first constraints on the $f$-mode frequency of the NSs in GW170817. In their work, they used a model for the gravitational waveform with an explicit dependence on the $f$-mode frequency which, in one of their analyses, was treated as an independent parameter, i.e. it was not determined by the isotropic $f$-Love relation. For GW170817, we use the publicly-available samples for $x$ and $y$ from Pratten et al.~\cite{Pratten:2019sed}, obtained with quadrupolar tidal contributions and without using URs on the waveform (see Table~1 in~\cite{Pratten:2019sed}). For future prospects, we use the samples for $x$ and $y$ from Williams et al.~\cite{Williams:2022vct}, for a GW170817-like event detected by a CE/ET network (see Fig. 9 in~\cite{Williams:2022vct}).

        \subsection{Bayesian Framework}

            \begin{figure*}
                \centering
                \includegraphics[width=\linewidth]{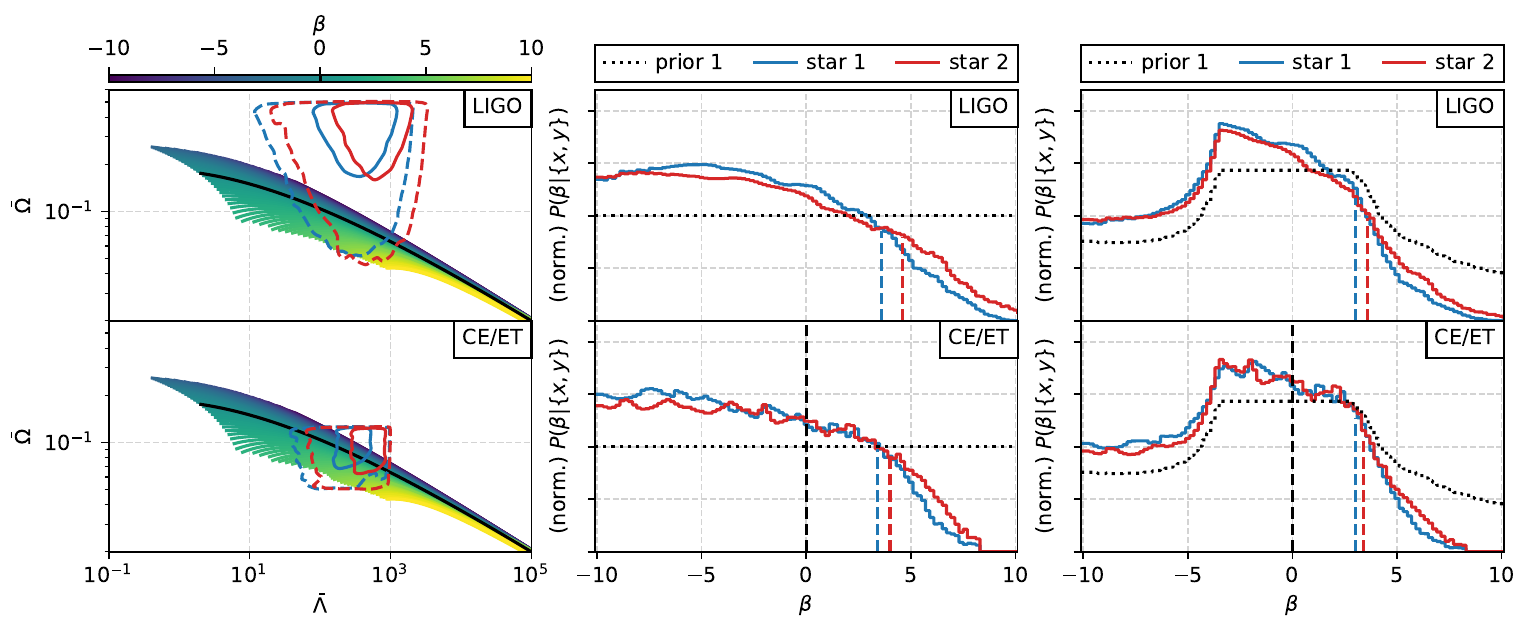}
                \caption{(Left)
                Posterior distributions on  $\bar{\Lambda}$ and $\bar{\Omega}$ for GW170817 from~\cite{Pratten:2019sed} (top) and for a GW170817-like event detected by a future CE/ET network from \cite{Williams:2022vct} (bottom). The solid and dashed lines represent the $50\%$ and $90\%$ credible contours, respectively. The contours for first and secondary stars ($M_{1} \geq M_{2}$) are in blue and red, respectively. We also show the anisotropic $f$-Love relations for various $\beta\in\left[-10,10\right]$, highlighting the isotropic $f$-Love relation in black. {\color{black} (Middle) Posterior distributions on $\beta$ for the primary and secondary stars, obtained with LIGO data (Top) and simulated CE/ET data (Bottom) while considering the uniform prior on $\beta$ (``prior 1'', see main text). The colored vertical dashed lines represent the $90\%$ upper bounds on $\beta$ and the black vertical dashed line indicates $\beta=0$, which should be recovered by our analysis since the simulated CE/ET data from~\cite{Williams:2022vct} was obtained for isotropic NSs. (Right) Same as middle, but using the physics-informed prior for $\beta$ (``prior 2'', see main text). For both choices of priors (see also Table~\ref{tab1}), we note that the 90$\%$ upper bounds for $\beta$ are similar for current LIGO data and simulated CE/ET data, although the physics-informed prior seems to give slightly lower bounds.} 
                } 
                \label{fig3}
            \end{figure*}

            Let us first review the Bayesian framework for computing a posterior distribution on $\beta$ from the inference of $\{x,y\}$. We follow Miller et al.~\cite{2020ApJ...888...12M}.
            Using the anisotropy-dependent, EOS-independent $f$-Love relations $y_{\rm UR}(x,\beta)$, we obtain the posterior $P$ on $\beta$ as
            \begin{align}
                P(\beta|\{x,y\})\propto p(\beta)\mathcal{L}(\{x,y\}|\beta), \label{posterior}
            \end{align}
            {\color{black} where $p(\hdots)$ is the prior and $\mathcal{L}(\hdots)$ is the likelihood, as we discuss below.}
            
            \subsubsection{Prior}
            
            {\color{black} We have two choices for the prior. First, we can choose a uniform prior, given by
            \begin{align}
                &p(\beta)=\mathcal{U}(\beta|\beta^{\rm min},\beta^{\rm max}),
            \end{align}
            namely a uniform distribution $\mathcal{U}$ for $\beta$ between $\beta^{\rm min}=-10$ and $\beta^{\rm max}=10$. We call this choice of prior ``prior 1''. Our second choice of prior takes into consideration the parameter space for physical models, which is determined by the conditions in Eqs.~\eqref{wec}$-$\eqref{causality}. One single model for an anisotropic star is characterized by its EOS ($p_{r}$~vs.~$\rho$ relation), central density $\rho_{0}$, and anisotropy parameter $\beta$. In our case, we have a three-dimensional grid of $5 \times 50 \times 50$ models, and these three numbers correspond to the different choices of EOS, $\rho_{0}$, and $\beta$, respectively. Thus, in this case, the prior on $\beta$ given the EOS is
            \begin{align}
                p(\beta|\textrm{EOS}) \propto \frac{N_{\rm physical}(\beta|\textrm{EOS})}{N_{\rm total}(\beta|\textrm{EOS})},\label{prior2}
            \end{align}
            where $N_{\rm physical}(\beta|\textrm{EOS}) \leq 50$ is the number of models that satisfy the conditions and $N_{\rm total}(\beta|\textrm{EOS})=50$. We use proportionality in Eq.~\eqref{prior2} because we need to normalize the expression according to
            \begin{align}
                \int p(\beta|\textrm{EOS}){\rm d}\beta = 1.
            \end{align}
            Finally, we marginalize over the EOSs, i.e.
            \begin{align}
                p(\beta)=\sum_{\rm EOS}p(\beta|\textrm{EOS})p(\textrm{EOS}),
                \label{prior_beta_1}
            \end{align}
            where, if no EOS is preferred a priori, $p(\textrm{EOS})=1/N_{\textrm{EOS}}$, where $N_{\textrm{EOS}}=5$ is the number of EOSs that we use\footnote{\color{black}One can further use recent astrophysical observations to assign weights to EOS models. For example, Biswas~\cite{Biswas:2021pvm} performed a Bayesian model selection of EOS models using GW, X-ray, and radio data; we tried using the Bayes factors for the EOSs reported in the study to construct a new prior for the EOS models considered here and found that the results are similar to the case for a uniform prior on the EOS. We do not impose the above weights to the EOS models in our main analysis as the analysis in~\cite{Biswas:2021pvm} assumes isotropic NSs, which may introduce a bias to our analysis.}. We call this choice of prior ``prior 2.''} 

            %\ky{Do you mean a uniform prior on EOS?} \vg{Yes! I added this part.}

            %If we consider the Bayes factors $\mathcal{B}^{\textrm{EOS}}_{\textrm{MPA1}}$, i.e. those relative to MPA1 (see right panel of their Figure 2), we have:
            %\begin{align}
                %p(\textrm{EOS})\propto\mathcal{B}^{\textrm{EOS}}_{\textrm{MPA1}},
                %\label{prior_eos}
            %\end{align}
            %where we use proportionality because we need to normalize the expression as
            %\begin{align}
                %\sum_{\textrm{EOS}}p(\textrm{EOS})=1.
            %\end{align}
            %Using the EOS prior in Eq.~\eqref{prior_eos} in Eq.~\eqref{prior_beta_1}, we obtain a physics- and data-informed prior on $\beta$.

            \subsubsection{Likelihood}
            
            The likelihood of the data (samples of $x$ and $y$), given the model (values of $x$ and $y$, given $\beta$) is:
            \begin{align}
                &\mathcal{L}(\{x,y\}|\beta)=\iint P_{\rm GW}(x,y)P_{\rm UR}(x,y|\beta)\textrm{d}x\textrm{d}y, \label{likelihood}
            \end{align}
            %\vl{question about the notation: On the right hand side, $x, y$ are dummy variables and integrated out. How would the likelihood on the left depend on $x$ and $y$? Is it better to write it as $\mathcal{L}(d_{\rm GW}| \beta)$, and define $d_{\rm GW}$ as the data from the GW analysis? Same for Eq. (3.10).} \vg{agreed!}
            which is an overlap between the distribution of the data $P_{\rm GW}(x,y)$ and the distribution of the model $P_{\rm UR}(x,y|\beta)$. The subscript ``GW'' denotes that the distribution originates from GW inference, while the subscript ``UR'' denotes that the distribution comes from the URs with anisotropy, which represents our model. We can write the latter as
            \begin{align}
                P_{\rm UR}(x,y|\beta)=P_{\rm UR}(y|x,\beta)P_{\rm UR}(x|\beta), \label{P_UR}
            \end{align}
            for
            \begin{align}
                &P_{\rm UR}(y|x,\beta)=\mathcal{N}(y|y_{\rm UR}(x,\beta),s_{\rm UR}^{2}(\beta)), \label{gaussian} \\
                &P_{\rm UR}(x|\beta)=\mathcal{U}(x|{x}^{\rm min}_{\rm UR}(\beta),{x}^{\rm max}_{\rm UR}(\beta)),
            \end{align}
            where $\mathcal N$ represents the normal distribution while ${x}^{\rm min}_{\rm UR}(\beta)$ and ${x}^{\rm max}_{\rm UR}(\beta)$ define the range of validity of the UR for each $\beta$. These values depend on the choice of EOSs to ``calibrate'' the URs, i.e. ${x}^{\rm min}_{\rm UR}(\beta)$ is the minimum of a set of
            \begin{align}
                x^{\rm min}_{\rm EOS}(\beta)=x_{\rm EOS}(\beta,M=M_{\rm max}),
            \end{align}
            that is computed for each EOS,  where $M_{\rm max}$ is the maximum mass\footnote{Note that, depending on $\beta$, the maximum mass cannot be reached because of the conditions in Eqs.~\eqref{wec}$-$\eqref{causality}. In these cases, $M_{\rm max}$ is the maximum value for the mass achieved by physically viable models.}. Similarly, ${x}^{\rm max}_{\rm UR}(\beta)$ is the maximum of a set of
            \begin{align}
                x^{\rm max}_{\rm EOS}(\beta)=x_{\rm EOS}(\beta,M=0.5\textrm{ M$_{\odot}$}).
            \end{align}
            We can interpret Eq.~\eqref{P_UR} in the following way. The probability density assigned to the pair $\{x,y\}$, given $\beta$, for our model, depends on: the probability density assigned to $y$, given $x$ and $\beta$, which can be approximated as a Gaussian with mean and variance as in Eqs.~\eqref{fit} and~\eqref{var}, respectively; the probability density assigned to $x$, given $\beta$, which is a uniform distribution defined on the range of validity of the UR.
            
            The advantage of performing the statistical inference using URs is that, given their weak EOS dependence, we have $s^{2}_{\rm UR}(\beta)\approx 0$ (which is indeed the case, see Table~\ref{tab1}). Then, from Eq.~\eqref{gaussian}, we have:
            \begin{align}
                P_{\rm UR}(y|x,\beta)\approx\delta(y-{y}_{\rm UR}(x,\beta)),
            \end{align}
            and thus we can write Eq.~\eqref{likelihood} as
            \begin{align}
                &\mathcal{L}(\{x,y\}|\beta)\propto\int P_{\rm GW}(x,y_{\rm UR}(x,\beta))\textrm{d}x.\label{likelihood_new}
            \end{align}
            which is faster to compute, when compared to Eq.~\eqref{likelihood}. This integral is computed between ${x}^{\rm min}_{\rm UR}(\beta)$ and ${x}^{\rm max}_{\rm UR}(\beta)$. Thus, we can compute the posterior in Eq.~\eqref{posterior} for our interval of $\beta$ and obtain credible regions for this parameter. 
            
            %\vl{Should we add lower and upper limits for the integration of $x$ directly in Eq. 3.10 or mention it in the text?}

        \subsection{Constraints on Pressure Anisotropy}

            \begin{table*}
                \centering
                \begin{ruledtabular}
                    \begin{tabular}{ccc}
                        data used & 90$\%$ upper bound on $\beta$ {\color{black}(prior 1)} & 90$\%$ upper bound on $\beta$ {\color{black}(prior 2)} \\
                        \hline
                        LIGO GW170817-1~\cite{Pratten:2019sed} & {\color{black} 3.6} & {\color{black} 3.0} \\
                        LIGO GW170817-2~\cite{Pratten:2019sed} & {\color{black} 4.6} & {\color{black} 3.6} \\
                        \hline
                        simulated CE/ET GW170817-like-1~\cite{Williams:2022vct} & {\color{black} 3.4} & {\color{black} 3.0} \\
                        simulated CE/ET GW170817-like-2~\cite{Williams:2022vct} & {\color{black} 4.0} & {\color{black} 3.4} \\
                    \end{tabular}
                    \caption{Results for the upper bounds on the anisotropy parameter $\beta$, using the inferences for GW170817 from Pratten et al.~\cite{Pratten:2019sed} and simulated data from Williams et al.~\cite{Williams:2022vct} (vertical dotted lines in the right panel of Fig.~\ref{fig3}), {\color{black} using the uniform prior (``prior 1'') and physics-informed prior (``prior 2'') on $\beta$. For both choices of priors, we obtained similar upper bounds for current LIGO data and simulated CE/ET data.}}
                    \label{tab2}
                \end{ruledtabular}
            \end{table*}
    
            % In Fig.~\ref{fig3}, we present our results for the posterior on $\beta$, using the inferences of $(x,y)$ for GW170817 from Pratten et al.~\cite{Pratten:2019sed} and for a GW170817-like event using a future CE/ET network from Williams et al.~\cite{Williams:2022vct}. 
            
            In the left panels of Fig.~\ref{fig3}, we present the contours for the distributions $P_{\rm GW}(x,y)$ for the two NSs in GW170817 from Pratten et al.~\cite{Pratten:2019sed} (top) and for NSs in a GW170817-like event detected by a future CE/ET network from Williams et al.~\cite{Williams:2022vct} (bottom). For comparison, we also show the anisotropic $f$-Love relations for various $\beta$ in the same panels.  For GW170817, note that the URs for $\beta \in [-10,10]$~\footnote{For each central energy density, we consider $\beta \in [-10,10]$ for $\Delta\beta=0.1$. Solutions whose values of $\beta$ violate the conditions in Eqs.~\eqref{wec}$-$\eqref{causality} are not considered.} are not contained within the 50\% credible region of the parameter space, while the relations are consistent with the 90\% credible region, except when $\beta$ takes a largely positive value. For the case of a simulated GW170817-like event detected by a future GW network, on the other hand, the URs are contained within the  50\% credible region of the GW inference for most values of $\beta$, while the situation is similar to the GW170817 case if we consider the 90\% credible region. 
            
            {\color{black} The middle and right panels of Fig.~\ref{fig3} show the posteriors on $\beta$ for the primary and secondary stars, $P_{1}(\beta|\{x,y\})$ and $P_{2}(\beta|\{x,y\})$, for both GW170817 data (top) and simulated GW170817-like data (bottom), considering the uniform (``prior 1'', middle) and physics-informed (``prior 2'', right) priors on $\beta$}. As expected, largely positive values of $\beta$ do not have much posterior support for both  the GW170817 data and the simulated data. We can put a 90\% credible upper bound on $\beta$, as indicated by {\color{black} colored} vertical dotted lines in the figure. We summarize in Table~\ref{tab2} the bounds on $\beta$ for the primary and secondary stars obtained with current (real) data and future (simulated) data {\color{black} for both choices of prior}. We stress that these bounds have minimal contamination from uncertainties in the EOS. 
            {\color{black} The bounds obtained with the physics-informed prior are slightly lower than the ones obtained with the uniform prior, and, interestingly, current and future bounds are very similar.} This similarity is mainly because the bulk of the distribution $P_{\rm GW}(x,y)$ for GW170817 has a large offset from the URs. Therefore, although the distributions are much wider for GW170817 than those for the simulated GW170817-like event for a CE/ET network, the boundary of the 90\% credible regions is comparable for the two cases in the regime where the contours cross the UR curves, making the 90\% credible bounds to be also comparable.

            In summary, we obtained similar constraints on the parameter $\beta$ from a single real and simulated event for a binary NS merger. However, we might be able to obtain tighter constraints by combining multiple events. Such an analysis is not trivial because $\beta$ is likely not to be a universal parameter among different sources, but rather a source-dependent parameter. Thus, we cannot simply multiply posteriors on $\beta$ to obtain joint constraints, although it could be possible to do a hierarchical analysis to constrain the distribution of $\beta$, as originally proposed in~\cite{Isi:2019asy} for tests of gravity with GWs.

            %\sout{i.e. the distribution of $\beta$ for different stars do not necessarily correspond to the same physics that could be attributed to $\beta$.} \ky{If the origin of anisotropy is e.g. magnetic field, then it comes from the same physical origin, but $\beta$ should depend on the magnetic field strength $B$ which depends on the source. I suggest to remove the last part that I striked out.}
            
            Besides that, since our phenomenological model assumes the simplest form that preserves the regularity of the background and perturbation equations (see discussion in~Sec.~\ref{subsec_aniso}), it is unclear how the bounds that we obtained on $\beta$ could be translated to more physically motivated models. For example, in~\cite{Dong:2024lte}, the authors found that the pressure anisotropy of a hybrid star with an elastic core can be fitted well with Eq. (1) of their paper, whose expression is more complicated than that in Eq.~\eqref{ansatz}, and has two free parameters instead of one. Thus, in order to draw conclusions about specific physics that causes pressure anisotropy in NSs, we need phenomenological models that can be mapped to such an origin of anisotropy.

            We point out that the main purpose of this study is to show how the universal $f$-Love relation for isotropic stars is affected by a simple phenomenological model for pressure anisotropy. We further show, in practice, how modified URs can be used to constrain fundamental physics, i.e. how measurements of the tidal deformability and the $f$-mode of NSs can constrain the anisotropy parameter in an EOS-independent way. Thus, if the $f$-Love relation is affected in a similar way for other anisotropy models that are more directly motivated from specific origins of anisotropy, the procedure developed here can be applied to constrain the physics behind pressure anisotropy in NSs with current and future GW observations.

            Finally, we mention that the reason for this anisotropy-induced non-universality of the $f$-Love relation is currently unknown. The pressure anisotropy gives rise to an extra term in the TOV equation (see Eq.~\eqref{eqpr} in Appendix~\ref{ap_Lf}) that acts like an additional force, which may not be captured by simply changing the EOS, although a more in-depth study is needed. Such a study may further provide hints about the origin of the EOS universality, which is also not yet fully understood.
            
    \section{Conclusions and Discussions}
    \label{sec4}

        We computed the tidal deformability~\cite{Yagi:2015hda} and $f$-mode frequency~\cite{Lau:2024oik} of anisotropic NSs using a quasi-local ansatz for the anisotropy (proposed in~\cite{Lau:2024oik}, inspired by~\cite{Horvat:2010xf}), described by a dimensionless parameter $\beta$. 
        We found that the $f$-Love relation for isotropic NSs depends on the anisotropy parameter $\beta$, although it remains EOS-insensitive for a fixed $\beta$. We took advantage of this anisotropy-dependent UR in our statistical approach to obtain upper bounds on $\beta$, using GW data from the analyses of real GW170817 data by LIGO~\cite{Pratten:2019sed} and simulated GW170817-like data by CE/ET~\cite{Williams:2022vct}, {\color{black} using a uniform and a physics-informed prior on $\beta$. The upper bounds on $\beta$ for the physics-informed prior are slightly lower than the ones obtained with the uniform prior, although very similar.} The bounds for real and simulated data are also very similar, and this similarity is more of a coincidence and it arises mainly because of the large offset in the most probable values of the tidal deformability and the $f$-mode frequency with GW170817 from the theoretical prediction (the $f$-Love relation). Although the estimated upper bounds are comparable, the network of third generation detectors would allow us to perform more precise tests of pressure anisotropy due to much smaller statistical errors in the two observable quantities. We stress that the bounds obtained here are EOS insensitive and our analysis explicitly shows how URs can be used in practice to test fundamental physics. 
       
        Various avenues exist for future work.
        For example, we used an effective model for the local anisotropy inside NSs that does not necessarily originate from a specific theory that describes anisotropic fluids. This brings in the issue of how the anisotropy responds to non-radial perturbations, since the quasi-local parameter is defined in the spherically symmetric configuration. Our current perturbation model, following \cite{Doneva:2012rd}, assumes the stress-energy tensor retains the same form as Eq.~\eqref{eqTmunu} in the non-radial configuration. This can be restrictive and may not accurately represent a generic anisotropic material without spherical symmetry. Thus, we could repeat the analysis with anisotropy models that are more physically motivated, such as the one proposed in~\cite{Cadogan:2024mcl,Cadogan:2024ohj,Cadogan:2024ywc} based on liquid crystals or consider elastic stars that are known to produce anisotropy~\cite{Karlovini:2002fc, Alho_2022, Dong:2024lte}. This approach requires one to first establish a perturbative framework for computing the tidal deformability and $f$-mode frequency under these anisotropy models.
        %as in e.g. elastic stars~\cite{Karlovini:2002fc}, where pressure anisotropy arises naturally from shear stresses. Thus, studying non-radial oscillations of such models constitutes one of the future directions of this work.  

        Let us also point out that the violation of the $f$-Love relation in NSs from the original one found for isotropic pressure can possibly bias the analysis of GW data if the gravitational waveform model assumes the relation to be valid. For example, the waveform model for binary NS mergers in~\cite{Abac:2023ujg} that was calibrated with numerical relativity simulations makes use of the original $f$-Love relation. Therefore, another extension of this work is to reanalyze the GW170817 data using the anisotropy-dependent $f$-Love relations, and infer the anisotropy parameter that best describes the data to check the consistency with the results presented here. 
        
        As a final remark, deviations from the isotropic $f$-Love relation can be caused by other factors other than anisotropy, like the mismodelling of the dynamical tide response in the inspiral signal.
        Recent studies show that the non-linear component of the dynamical tide can enhance the tidally induced phase shift in the late inspiral signal by $\sim 10\%$ \cite{Yu:2022fzw, Pitre:2025qdf}, and therefore need to be incorporated in the waveform for precise measurements of the $f$-mode. In addition, the previous models using the effective Love number approach ignore the effect of tidal spin, which can contribute to a phase shift up to the same order depending on the background spin of the NS \cite{Yu_2025}. These can contribute to the systematic error when inferring the $f$-mode frequency in the GW data. \\

        {\color{black}
        {\it Note added}. During the review of this manuscript, the work from Becerra et al.~\cite{Becerra:2025wno} was announced on arXiv. The authors studied the radial stability of anisotropic stars through their dynamical evolution with a fully non-linear relativistic code. Their findings for the modified H-model (Eq.~\eqref{ansatz}) agree with what we report in Appendix~\ref{apA}.
        }

    \begin{acknowledgments}
        The authors thank Geraint Pratten for providing the samples in Fig. 9 of Williams et al.~\cite{Williams:2022vct}. V.G. thanks Cole Miller for discussions. S.A. and K.Y. acknowledge support from NSF Grant \textcolor{black}{PHY-2309066 and} PHY-2339969. V.G., S.A. and K.Y. also acknowledge support from the Owens Family Foundation. S.Y.L. acknowledges support from Montana NASA EPSCoR Research Infrastructure Development under award No. 80NSSC22M0042.

    \end{acknowledgments}

    % \begin{appendices}
    % \counterwithin{equation}{section}

    \appendix
          
    \section{Radial stability of anisotropic neutron stars}
    \label{apA}

        \begin{figure*}
            \centering
            \includegraphics[width=\linewidth]{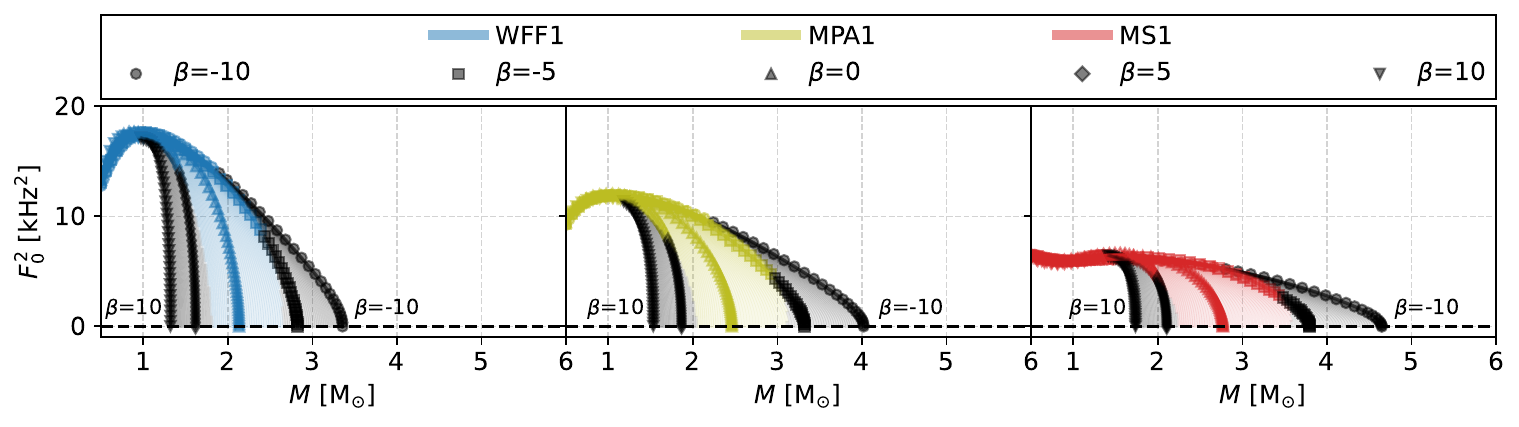}
            \caption{Squared frequency of the fundamental radial mode of anisotropic NSs for the anisotropy model in Eq.~\eqref{ansatz} for various $\beta$. Each panel shows the result for one EOS. Each sequence in grey stops at the maximum mass while we highlight physically-viable stellar configurations satisfying  the conditions in Eqs.~\eqref{wec}$-$\eqref{causality} in color. This result confirms that these models satisfy the maximum-mass criterion for the radial stability of anisotropic NSs, since $F^{2}_{0}(M)\rightarrow0$ as $M\rightarrow{M_{\rm max}}$. }
            \label{fig4}
        \end{figure*}

        In this Appendix, we study the radial stability of NSs with pressure anisotropy in Eq.~\eqref{ansatz}. Such a study can be done by considering radial perturbations of equilibrium solutions and check at what central density $\rho_{0}$ the square of the fundamental radial mode frequency $F^{2}_{0}$ crosses zero. For NSs with isotropic pressure, the critical energy density coincides with that for the maximum mass configuration. Namely, the star becomes radially unstable when $\partial{M}/\partial{\rho_{0}} < 0$ and this is called the \emph{maximum-mass criterion}.

        %   NSs have a maximum mass $M_{\rm max}$, and if $M(\rho_{0})>M_{\rm max}$, such that $\partial{M}/\partial{\rho_{0}} < 0$, they are radially unstable. This ``maximum-mass criterion'' is usually considered when obtaining stable solutions for NSs. However, a more robust criterion is to consider radial perturbations of equilibrium solutions and find the density such that the square of the fundamental radial mode frequency $F^{2}_{0}$ crosses zero. 
        
        % The energy density corresponding to the maximum mass of a NS, for a certain EOS, matches the one obtained by studying radial oscillations. 
        This criterion is not guaranteed to hold for relativistic stars other than NSs or those with anisotropic pressure. For example, hybrid stars are known to have extended branches of stability beyond their maximum mass~\cite{Lugones:2021zsg}. The radial stability of anisotropic NSs has been studied previously by, e.g. Pretel~\cite{Pretel:2020xuo}, and the author found that some of the phenomenological models for anisotropy (e.g. Bowers and Liang~\cite{Bowers:1974tgi}) also break this correspondence, thus making the maximum-mass criterion not reliable when determining radially stable solutions for anisotropic NSs.

        The radial stability of NS models described by the original ansatz by Horvat et al.~\cite{Horvat:2010xf} were studied in the same paper and also by Pretel~\cite{Pretel:2020xuo}. The studies found that the maximum-mass density matches with the critical density for stability obtained from radial oscillation calculations. In this work, we considered a modified version of the model in Horvat et al.~\cite{Horvat:2010xf}. Since the radial stability of NSs with such a model has not been studied before, we check the stability here for the first time.
        %Thus, since we are interested in computing observable quantities for anisotropic NSs, such as the tidal deformability and $f$-mode frequency, described by this model, we found it worth it to study their radial stability and either confirm or disprove that the maximum-mass criterion can be used for this ansatz.

        We follow Pretel~\cite{Pretel:2020xuo}, i.e. we consider adiabatic radial oscillations by perturbing the structure equations, maintaining spherical symmetry. We find the fundamental radial mode by solving the equations for the Lagrangian displacement and the Lagrangian perturbation on the pressure, under the appropriate boundary conditions. We show the results for three EOSs of different stiffness in Fig.~\ref{fig4}. We find that the maximum-mass criterion is valid for our anisotropy model, namely $\partial{M}/\partial{\rho_{0}} < 0$ if and only if $F^{2}(\rho_{0}) <0$, similar to NSs with isotropic pressure. We stress that even models that are radially stable can violate the energy conditions or causality (as in Eqs.~\eqref{wec}$-$\eqref{causality}). We highlight models that are physically viable in color in Fig.~\ref{fig4}. These are the stellar models that we considered in our analysis in Sec.~\ref{sec3}.

    \section{Computational framework for tidal deformability and non-radial oscillations of anisotropic neutron stars}
    \label{ap_Lf}
    \counterwithin{subsection}{section}

        In this Appendix, we review the equations that we need to solve to compute the tidal deformability and the fundamental mode of anisotropic stars. We begin by describing necessary equations for constructing spherically-symmetric stars.
        The stress-energy tensor for a spherically-symmetric  anisotropic fluid is given by
        \begin{align}
            T_{\mu\nu}=\rho u_{\mu}u_{\nu}+p_{r}\left(g_{\mu\nu}+u_{\mu}u_{\nu}\right)-\sigma\left(g_{\mu\nu}+u_{\mu}u_{\nu}-k_{\mu}k_{\nu}\right), \label{eqTmunu}
        \end{align}
        where $u_{\mu}$ is the fluid four-velocity satisfying $u_{\mu}u^{\mu}=-1$, and $k_{\mu}$ is the unit, radial vector satisfying $k_{\mu}k^{\mu}=1$ and $k_{\mu}u^{\mu}=0$. The line-element for a static, spherically symmetric spacetime is
        \begin{align}
            {\rm d}s^{2}=g_{\mu\nu}{\rm d}x^{\mu}{\rm d}x^{\nu}=-{\rm e}^{\nu}{\rm d}t^{2}+{\rm e}^{\lambda}{\rm d}r^{2}+r^{2}\left({\rm d}\theta^{2}+\sin^{2}{\theta}{\rm d}\phi^{2}\right),
        \end{align}
        where $\nu(r)$ and $\lambda(r)$ are functions of $r$, and we define the mass function as
        \begin{align}
            m(r)\equiv\frac{r}{2}\left(1-{\rm e}^{\lambda}\right).
        \end{align}
        We substitute the above stress-energy tensor and the line element to the Einstein equations and the stress-energy conservation:
        \begin{align}
            &G_{\mu\nu}-8\pi T_{\mu\nu}=0, \\
            &\nabla^{\mu}T_{\mu\nu}=0.
        \end{align}
        We then find the following equations for the background spherically-symmetric spacetime:
        \begin{align}
            &\frac{{\rm d}m}{{\rm d}r}=4{\pi}r^{2}\rho,\label{eqm}\\
            &\frac{{\rm d}\nu}{{\rm d}r}=\frac{2{\rm e}^{\lambda}}{r^{2}}\left(m+4{\pi}r^{3}p_{r}\right),\label{eqnu}\\
            &\frac{{\rm d}p_{r}}{{\rm d}r}=-\frac{\left(\rho+p_{r}\right)}{2}\frac{{\rm d}\nu}{{\rm d}r}-\frac{2\sigma}{r}. \label{eqpr}
        \end{align}
        To solve the above equations, we impose the boundary conditions around $r=0$ as
        \begin{align}
            &m(r)=\frac{4\pi}{3}\rho^{(0)}r^{3}+\mathcal{O}\left(r^{5}\right),\\
            &\nu(r)=\nu^{(0)}+\frac{1}{2}\nu^{(2)}r^{2}+\mathcal{O}\left(r^{4}\right),\\
            &p_{r}(r)=p_{r}^{(0)}+\frac{1}{2}p_{r}^{(2)}r^{2}+\mathcal{O}\left(r^{4}\right),\\
            &\rho(r)=\rho^{(0)}+\frac{1}{2}\rho^{(2)}r^{2}+\mathcal{O}\left(r^{4}\right),\\
            &\sigma(r)=\frac{1}{2}\sigma^{(2)}r^{2}+\mathcal{O}\left(r^{4}\right),
        \end{align}
        with the coefficients
        \begin{align}
            &\nu^{(2)}=\frac{8\pi}{3}\left(3p_{r}^{(0)}+\rho^{(0)}\right)r^{2},\\
            &p_{r}^{(2)}=-\frac{4\pi}{3}\left(\rho^{(0)}+p_{r}^{(0)}\right)\left(\rho^{(0)}+3p_{r}^{(0)}\right){\color{black}-2\sigma^{(2)}},\\
            &\rho^{(2)}=\frac{p_{r}^{(2)}\left(\rho^{(0)}+p_{r}^{(0)}\right)}{\gamma^{(0)}p_{r}^{(0)}},
        \end{align}
        where the subscript $(n)$ represents the order of the expansion of the function in small $r$. $\gamma$ is the adiabatic index defined as
        {\color{black}
        \begin{align}
            \gamma\equiv \frac{\rho+p_{r}}{p_{r}}c^{2}_{s,r},
        \end{align}
        where $c^{2}_{s,r}$ is defined in Eq.~\eqref{sound_speed}}, and, for the anisotropy ansatz in Eq.~\eqref{ansatz},
        \begin{align}
            \sigma^{(2)}=0.
        \end{align}
    
        Having these background equations at hand, we will next introduce perturbations to the metric ($\delta g_{\mu\nu}$) and matter fields ($\delta\rho$, $\delta p_{r}$, and $\delta\sigma$) to derive key equations for computing the tidal deformability and $f$-mode frequency from the perturbed Einstein equations and stress-energy conservation condition:
        \begin{align}
            &\delta(G_{\mu\nu}-8\pi T_{\mu\nu})=0, \label{dpe1}\\
            &\delta(\nabla^{\mu}T_{\mu\nu})=0. \label{dpe2}
        \end{align}
        Below, we will review the calculation for the two quantities separately.

        \subsection{Tidal deformability}
        \label{ap_L}

            First, we focus on static tidal perturbations to compute the tidal deformability.
            Following Hinderer~\cite{Hinderer:2007mb} and Yagi \& Yunes~\cite{Yagi:2015hda}, the line element for static perturbations on the metric in the Regge-Wheeler gauge~\cite{Regge:1957td}, is given by
            \begin{align}
                \delta g_{\mu\nu}{\rm d}x^{\mu}{\rm d}x^{\nu}=-\sum_{\ell,m}\left[{\rm e}^{\nu}H_{0}{\rm d}t^{2}+{\rm e}^{\lambda}H_{2}{\rm d}r^{2}+r^{2}K\left({\rm d}\theta^{2}+\sin^{2}{\theta}{\rm d}\phi^{2}\right)\right]Y_{{\ell}m},
            \end{align}
            where $H_{0}(r)$, $H_{2}(r)$, and $K(r)$ are functions of $r$ and $Y_{{\ell}m}(\theta,\phi)$ are the spherical harmonics. From Eqs.~\eqref{dpe1} and~\eqref{dpe2}, we obtain:
            \begin{align}
            \label{eq:H0_eq}
                &\frac{{\rm d}^{2}H_{0}}{{\rm d}r^{2}}+\left\{\frac{2}{r}+2{\rm e}^{\lambda}\left[\frac{m}{r^{2}}+2\pi{r}(p_r - \rho)\right]\right\}\frac{{\rm d}H_{0}}{{\rm d}r}-\left\{{\rm e}^{\lambda}\left[\frac{\ell\left(\ell+1\right)}{r^{2}}-16\pi\left(2p_{r}+\rho-\sigma\right)\right]+\left(\frac{{\rm d}\nu}{{\rm d}r}\right)^{2}\right\}H_{0}\nonumber\\&-8\pi{\rm e}^{\lambda}\left(1+\frac{\rho+p_{r}}{\gamma{p_{r}}}\right)\delta{p_{r}}=0,\\
              \label{eq:p1_eq}
                &\frac{{\rm d}\left(\delta{p_{r}}\right)}{{\rm d}r}+\frac{1}{2r}\left[{\rm e}^{\lambda}\left(1+\frac{\rho+p_{r}}{\gamma{p_{r}}}\right)\left(8\pi{r^{2}}p_{r}+1\right)-\frac{\rho+p_{r}}{\gamma{p_{r}}}+3\right]\delta{p_{r}}+\left(\frac{\rho+p_{r}}{2}-\sigma\right)\frac{{\rm d}H_{0}}{{\rm d}r}\nonumber\\
                &+\frac{1}{r}\left[\rho+p_{r}+\sigma-{\rm e}^{\lambda}\sigma\left(8\pi{r^{2}}p_{r}+1\right)\right]H_{0}=0.
            \end{align}
            We have checked that the above equations are consistent with Eqs.~(22)--(26) in~\cite{Yagi:2015hda} with the identification $\omega_1 = 0$, $h_2 = H_0/2$ and $\xi_2 = - ({\rm d}p_r/{\rm d}r)\delta p_r$. Note that the above equations are a set of a 3rd-order differential equations and is different from the 2nd-order system that was originally proposed in~\cite{Biswas:2019gkw} for anisotropic NSs following closely the isotropic case in~\cite{Hinderer:2007mb}. We will show more detailed derivation of Eqs.~\eqref{eq:H0_eq} and~\eqref{eq:p1_eq} and clarify the difference in the two formulations for static tidal perturbations on anisotropic NSs in a forthcoming paper~\cite{yagi2025}.
             
            The $\ell=2$ relativistic electric tidal Love number is obtained from $H_0$ and its derivative at the surface as
            \begin{align}
                k_{2}&=\frac{8}{5}C^{5}\left(1-2C\right)^{2}\left[2+2C\left(y_R-1\right)-y_R\right] \nonumber \\
                &\times\{2C\left[6-3y_R+3C\left(5y_R-8\right)\right]+4C^{3}\left[13-11y_R+C\left(3y_R-2\right)+2C^{2}(1+y_R)\right] \nonumber \\
                &+3(1-2C)^{2}\left[2-y_R+2C(y_R-1)\right]{\rm ln}\left(1-2C\right)\}^{-1},\label{eqk2}
            \end{align}
            where:
            \begin{align}
                y_R=\frac{r}{H_{0}}\frac{{\rm d}H_{0}}{{\rm d}r}\bigg|_{r=R}.
            \end{align}
            To find $y_R$, we solve Eqs.~\eqref{eq:H0_eq} and~\eqref{eq:p1_eq} under the boundary condition near $r=0$: 
            \begin{align}
                &H_{0}=\frac{1}{2}H^{(2)}_{0}r^{2}+\mathcal{O}(r^{4}),\\
                &\delta{p_{r}}=\frac{1}{2}\delta{p_{r}^{(2)}}r^{2}+\mathcal{O}(r^{4}),\\
                &\delta{\sigma}=\frac{1}{2}\delta{\sigma^{(2)}}r^{2}+\mathcal{O}(r^{4}),
            \end{align}
            for the coefficients
            \begin{align}
                \delta{p_{r}^{(2)}}=\delta{\sigma^{(2)}}=\frac{1}{2}H^{(2)}_{0}\left(\rho^{(0)}+p_{r}^{(0)}\right),
            \end{align}
            where $H^{(2)}_{0}$ is an arbitrary constant.

        \subsection{Non-radial oscillations}
        \label{ap_f}

            We next review the formulation for computing non-radial oscillations for anisotropic NSs, which requires us to introduce dynamical perturbations.
            Following Lindblom \& Detweiler~\cite{Lindblom:1983ps,Detweiler:1985zz}, the line element for dynamical perturbations on the metric in the Regge-Wheeler gauge~\cite{Regge:1957td}, is given by
            \begin{align}
                \delta{g_{\mu\nu}}{\rm d}x^{\mu}{\rm d}x^{\nu}=-\sum_{\ell,m}\left[{\rm e}^{\nu}{\bar{H}}_{0}{\rm d}t^{2}+2{\rm i}{\omega}r{\bar{H}}_{1}{\rm d}t{\rm d}r+{\rm e}^{\lambda}{\bar{H}}_{2}{\rm d}r^{2}+r^{2}\bar{K}({\rm d}\theta^{2}+\sin^{2}{\theta}{\rm d}\phi^{2})\right]r^{\ell}{\rm e}^{{\rm i}{\omega}t}Y_{{\ell}m},
            \end{align}
            where ${\bar{H}}_{0}(r)$, ${\bar{H}}_{1}(r)$, ${\bar{H}}_{2}(r)$, and $\bar{K}(r)$ are functions of $r$. $\delta$ on the metric perturbation refers to the Eulerian perturbation, which is related to the Lagrangian perturbation $\Delta$ via ${\Delta}=\delta+\mathcal{L}_{\xi}$, where $\mathcal{L}_{\xi}$ is the Lie derivative along the Lagrangian displacement vector $\xi^\mu$ given by
            \begin{align}
                \xi^{\mu}\partial_{\mu}=\sum_{{\ell}m}r^{\ell-1}{\rm e}^{{\rm i}{\omega}t}\left[{\rm e}^{-\lambda/2}\bar{W}Y_{{\ell}m}\partial_{r}-\frac{\bar{V}}{r}\left(\frac{\partial{Y_{{\ell}m}}}{\partial\theta}\partial_{\theta}+\frac{1}{\sin^{2}{\theta}}\frac{\partial{Y_{{\ell}m}}}{\partial\phi}\partial_{\phi}\right)\right],
            \end{align}
            with $\bar{V}(r)$ and $\bar{W}(r)$ representing some functions of $r$. 
            %The relation between Lagrangian ($\Delta$) and Eulerian ($\delta$) perturbation is ${\Delta}=\delta+\mathcal{L}_{\xi}$, where $\mathcal{L}_{\xi}$ is the Lie derivative along $\xi$. 
            The Lagrangian perturbation on the radial pressure is related to the one on the density by:
            \begin{align}
                {\Delta}p_{r}=\frac{{\gamma}p_{r}}{\rho+p_{r}}\Delta\rho.
            \end{align}
         
            From Eqs.~\eqref{dpe1} and~\eqref{dpe2}, we obtain the key equations to determine the dynamical perturbation variables:
            \begin{align}
                \frac{{\rm d}{\bar{H}}_{1}}{{\rm d}r}&=\left[4\pi\left(\rho-p_{r}\right){\rm e}^{\lambda}r-\frac{2m{\rm e}^{\lambda}}{r^{2}}-\frac{\ell+1}{r}\right]{\bar{H}}_{1}+\frac{{\rm e}^{\lambda}}{r}\bar{K}+\frac{{\rm e}^{\lambda}}{r}{\bar{H}}_{0}-\frac{16\pi\left(\rho+p_{r}\right){\rm e}^{\lambda}\left(1-\bar{\sigma}\right)}{r}\bar{V},\label{eqH1}\\
                \frac{{\rm d}\bar{K}}{{\rm d}r}&=\frac{\ell\left(\ell+1\right)}{2r}{\bar{H}}_{1}+\left(\frac{1}{2}\frac{{\rm d}\nu}{{\rm d}r}-\frac{\ell+1}{r}\right)\bar{K}+\frac{1}{r}{\bar{H}}_{0}-\frac{8\pi\left(\rho+p_{r}\right){\rm e}^{\lambda/2}}{r}\bar{W},\label{eqK}\\
                \frac{{\rm d}\bar{W}}{{\rm d}r}&=r{\rm e}^{\lambda/2}\left(1-\bar{\sigma}\right)\bar{K}-\left(\frac{\ell+1}{r}-\frac{2\bar{\sigma}}{r}\right)\bar{W}+\frac{r{\rm e}^{\left(\lambda-\nu\right)/2}}{{\gamma}p_{r}}\bar{X}+\frac{r{\rm e}^{\lambda/2}}{2}{\bar{H}}_{0}-\frac{\ell\left(\ell+1\right){\rm e}^{\lambda/2}\left(1-\bar{\sigma}\right)}{r}\bar{V},\label{eqW}\\
                \frac{{\rm d}\bar{X}}{{\rm d}r}&=\frac{\left(\rho+p_{r}\right){\rm e}^{\nu/2}}{2}\left[r\omega^{2}{\rm e}^{-\nu}+\frac{\ell\left(\ell+1\right)\left(1-2\bar{\sigma}\right)}{2r}\right]{\bar{H}}_{1}+\frac{\left(\rho+p_{r}\right){\rm e}^{\nu/2}}{2}\left[\left(\frac{3}{2}-2\bar{\sigma}\right)\frac{{\rm d}\nu}{{\rm d}r}-\frac{\left(1-6\bar{\sigma}\right)}{r}-\frac{4{\bar{\sigma}}^{2}}{r}\right]\bar{K}\nonumber\\
                &-\frac{\left(\rho+p_{r}\right){\rm e}^{\left(\lambda+\nu\right)/2}}{r}\left[4\pi\left(\rho+p_{r}\right)+\omega^{2}{\rm e}^{-\nu}-F\right]\bar{W}-\frac{1}{r}\left(\ell-\frac{2\left(\rho+p_{r}\right)\bar{\sigma}}{{\gamma}r}\right)\bar{X}\nonumber\\
                &+\frac{\left(\rho+p_{r}\right){\rm e}^{\nu/2}}{2}\left(\frac{1}{r}-\frac{1}{2}\frac{{\rm d}\nu}{{\rm d}r}\right){\bar{H}}_{0}+\frac{\ell\left(\ell+1\right){\rm e}^{\nu/2}}{r^{2}}\frac{{\rm d}p_{r}}{{\rm d}r}\left(1-\bar{\sigma}\right)\bar{V}+\frac{2{\rm e}^{\nu/2}}{r}\bar{S}\label{eqX}.
            \end{align}
            Here $\bar{\sigma}=\sigma/\left(\rho+p_{r}\right)$, 
            \begin{align}
                F={\rm e}^{-\lambda/2}\left\{\frac{r^{2}}{2}\frac{{\rm d}}{{\rm d}r}\left(\frac{{\rm e}^{-\lambda/2}}{r^{2}}\frac{{\rm d}\nu}{{\rm d}r}\right)-{\rm e}^{-\lambda/2}\left[\left(\frac{6}{r^{2}}-\frac{2}{r}\frac{{\rm d}\nu}{{\rm d}r}\right)\bar{\sigma}-\frac{2r}{r^{2}\left(\rho+p_{r}\right)}\frac{{\rm d}\sigma}{{\rm d}r}-\frac{4{\bar{\sigma}}^{2}}{r^{2}}\right]\right\},
            \end{align}
            the function $\bar{X}$ is related to the Lagrangian perturbation on the radial pressure by
            \begin{align}
                \sum_{\ell,m}r^{\ell}\bar{X}Y_{{\ell}m}{\rm e}^{{\rm i}{\omega}t}=-{\rm e}^{\nu/2}{\Delta}p_{r},
            \end{align}
            and the function $\bar{S}$ is related to the Eulerian perturbation on the anisotropy by
            \begin{align}
                \sum_{\ell,m}r^{\ell}\bar{S}Y_{{\ell}m}{\rm e}^{{\rm i}{\omega}t}=\delta\sigma.
            \end{align}
            $\bar{S}$ is related to the other perturbation functions by
            \begin{align}
                \bar{S}&=-\left\{\left(\frac{\partial{\sigma}}{\partial{p_{r}}}\right)+\left(\rho+p_{r}\right)\left[A\left(\frac{{\rm d}p_{r}}{{\rm d}r}\right)^{-1}+\frac{1}{{\gamma}p_{r}}\right]\left(\frac{\partial{\sigma}}{\partial{\rho}}\right)\right\}\frac{{\rm e}^{-\lambda/2}}{r}\frac{{\rm d}p_{r}}{{\rm d}r}\bar{W}\nonumber\\
                &-\left[\left(\frac{\partial{\sigma}}{\partial{p_{r}}}\right)+\frac{\left(\rho+p_{r}\right)}{{\gamma}p_{r}}\left(\frac{\partial{\sigma}}{\partial{\rho}}\right)\right]{\rm e}^{-\nu/2}\bar{X}-\left(\frac{\partial{\sigma}}{\partial{\mu}}\right){\rm e}^{-\lambda}{\bar{H}}_{0},\label{S}
            \end{align}
            where $A$ is the Schwarzschild discriminant, that we set to zero. We assume the Eulerian perturbation on the anisotropy to be~\cite{Yazadjiev:2011ks}:
            \begin{align}
                \delta\sigma=\frac{\partial{\sigma}}{\partial{p_{r}}}\delta{p_{r}}+\frac{\partial{\sigma}}{\partial{\rho}}\delta{\rho}+\frac{\partial{\sigma}}{\partial{\mu}}\delta{\mu}. \label{eq_dsigma}
            \end{align} 
            Further, the functions ${\bar{H}}_{0}$ and $\bar{V}$ are determined by the following algebraic relations:
            \begin{align}
                \left[3m+\frac{\left(\ell-1\right)\left(\ell+2\right)r}{2}+4{\pi}r^{3}p_{r}\right]{\bar{H}}_{0}&=8{\pi}r^{3}{\rm e}^{-\nu/2}\bar{X}-\left[\frac{\ell\left(\ell+1\right)\left(m+4{\pi}r^{3}p_{r}\right)}{2}-\omega^{2}r^{3}{\rm e}^{-\left(\lambda+\nu\right)}\right]{\bar{H}}_{1}\nonumber\\
                &+\left[\frac{\left(\ell-1\right)\left(\ell+2\right)r}{2}-\omega^{2}r^{3}{\rm e}^{-\nu}-\frac{{\rm e}^{\lambda}\left(m+4{\pi}r^{3}p_{r}\right)\left(3m-r+4{\pi}r^{3}p_{r}\right)}{r}\right]\bar{K}\nonumber\\
                &-16{\pi}r{\rm e}^{-\lambda/2}\left(\rho+p_{r}\right)\bar{\sigma}\bar{W},\label{H0}\\
                \omega^{2}\left(\rho+p_{r}\right){\rm e}^{-\nu/2}\left(1-\bar{\sigma}\right)\bar{V}&=\bar{X}-\frac{\left(\rho+p_{r}\right){\rm e}^{\nu/2}}{2}{\bar{H}}_{0}-\frac{1}{r}\frac{{\rm d}p_{r}}{{\rm d}r}{\rm e}^{\left(\nu-\lambda\right)/2}\bar{W}-{\rm e}^{\nu/2}\bar{S}. \label{V}
            \end{align}
          
            We follow the procedure outlined in Lindblom \& Detweiler~\cite{Lindblom:1983ps,Detweiler:1985zz} to solve Eqs.~\eqref{eqH1}$-$\eqref{eqX} inside the star ($r<R$). To do so, we impose boundary conditions at $r=0$ and $r=R$. Regarding the former, we require the solutions to be regular at the center, thus imposing the boundary conditions
            \begin{align}
                &{\bar{H}}_{1}={\bar{H}}^{(0)}_{1}+\frac{1}{2}{\bar{H}}^{(2)}_{1}r^{2}+\mathcal{O}(r^{4}),\label{H1exp}\\
                &\bar{K}=\bar{K}^{(0)}+\frac{1}{2}\bar{K}^{(2)}r^{2}+\mathcal{O}(r^{4}),\\
                &\bar{W}=\bar{W}^{(0)}+\frac{1}{2}\bar{W}^{(2)}r^{2}+\mathcal{O}(r^{4}),\\
                &\bar{X}=\bar{X}^{(0)}+\frac{1}{2}\bar{X}^{(2)}r^{2}+\mathcal{O}(r^{4}),\\
                &{\bar{H}}_{0}={\bar{H}}^{(0)}_{0}+\frac{1}{2}{\bar{H}}^{(2)}_{0}r^{2}+\mathcal{O}(r^{4}),\\
                &\bar{V}=\bar{V}^{(0)}+\frac{1}{2}\bar{V}^{(2)}r^{2}+\mathcal{O}(r^{4}),\\
                &\bar{S}=\bar{S}^{(0)}+\frac{1}{2}\bar{S}^{(2)}r^{2}+\mathcal{O}(r^{4})\label{Sexp},
            \end{align}
            where
            \begin{align}
                &{\bar{H}}^{(0)}_{1}=\frac{16\pi}{\ell\left(\ell+1\right)}\left(\rho^{(0)}+p^{(0)}_{r}\right)\bar{W}^{(0)}+\frac{2}{\ell+1}\bar{K}^{(0)},\\
                &\bar{X}^{(0)}=\left(\rho^{(0)}+p^{(0)}_{r}\right){\rm e}^{\nu^{(0)}/2}\left\{\left[\frac{4\pi}{3}\left(\rho^{(0)}+3p^{(0)}_{r}\right)+\frac{2\sigma^{(2)}}{\rho^{(0)}+p^{(0)}_{r}}-\frac{\omega^{2}{\rm e}^{-\nu^{(0)}}}{\ell}\right]+\frac{\bar{K}^{(0)}}{2}\right\},\\
                &\bar{S}^{(0)}=0, \label{eq_S_regularity}
            \end{align}
            To find two linearly-independent solutions, from $r=0$ to $r=R/2$, we further consider
            \begin{align}
                &\bar{W}^{(0)}=1,\\
                &\bar{K}^{(0)}=\pm\left(\rho^{(0)}+p^{(0)}_{r}\right).
            \end{align}
            The coefficients ${\bar{H}}^{(0)}_{0}$ and $\bar{V}^{(0)}$ can be obtained from Eqs.~\eqref{H0} and~\eqref{V}, and $\bar{S}^{(0)}$ is obtained from the regularity conditions in Eq.~\eqref{reg_con_2}. We omit the second-order coefficients in Eqs.~\eqref{H1exp}$-$\eqref{Sexp} for brevity. On the other hand, at $r=R$, the Lagrangian perturbation of the pressure vanishes, i.e. $\bar{X}(R)=0$. To obtain three linearly-independent solutions from $r=R$ to $r=R/2$, we consider the combinations $\left\{1,0,0\right\}$, $\left\{0,1,0\right\}$, and $\left\{0,0,1\right\}$ for $\left\{{\bar{H}}_{1}(R),\bar{K}(R),\bar{W}(R)\right\}$. With two linearly-independent solutions from $r=0$ to $r=R/2$ and three linearly-independent solutions from $r=R$ to $r=R/2$, we obtain the full solutions for ${\bar{H}}_{1}$, $\bar{K}$, $\bar{W}$, and $\bar{X}$ while keeping one of the coefficients for the linear combinations free. We obtain  $\bar{S}$, ${\bar{H}}_{0}$, and $\bar{V}$ through Eqs.~\eqref{S},~\eqref{H0}, and~\eqref{V}, respectively.
    
            We follow the standard procedure in Lindblom \& Detweiler~\cite{1983ApJS...53...73L} to solve the perturbation equations outside the star ($r>R$), and also consider the corrections in~\cite{2011ChPhB..20d0401L}. Namely, we match the solutions for ${\bar{H}}_{0}$ and ${\bar{K}}$ with the Zerilli function and its derivative at the surface ($r=R$), then we solve the Zerilli equation up to a large radius (usually $r=25/{\rm Re}(\omega)$, for $\omega$ being a guess for the complex frequency), and match it to the asymptotic solution of the Zerilli equation for an outgoing wave. We find the complex frequency that corresponds to an outgoing wave solution at large $r$ by using a shooting method.

    %\end{appendices}
            Lastly, we also explain the difference between our formulation with the one independently derived in \cite{Mondal:2023wwo}. We find that there are two major issues within their derivation, causing their pulsation equations to deviate from ours (compare Eqs.~\eqref{eqW}, \eqref{eqX} and \eqref{V} of this paper with the equations of the same dependent variables in \cite{Mondal:2023wwo}). First, the local energy conservation law (Eq.~(33) of \cite{Mondal:2023wwo}) is inconsistent with the law of thermodynamics of an anisotropic medium. Since the stress is locally anisotropic, the change in energy density should be directionally dependent (see Eq.~(A3) of \cite{Lau:2024oik}). As a result, their formulation does not reduce to the Cowling limit in \cite{Doneva:2012rd}. Second, the anisotropy model employed in \cite{Mondal:2023wwo, Mondal:2025ixk} causes the solution to become irregular at the stellar center. This is due to the non-zero $\partial \sigma/ \partial \mu$ term in the assumed form of $\delta \sigma$ (Eq.~\eqref{eq_dsigma}). As seen in Eq.~\eqref{eq_S_regularity}, this violates the regularity conditions.

    \bibliography{references}

\end{document}